\documentclass[aps,prd,groupedaddress,showpacs,twocolumn,eqsecnum]{revtex4}
\usepackage{graphicx}% Include figure files
\usepackage{dcolumn}% Align table columns on decimal point
\usepackage{bm}% bold math
\begin{document}

\title{Confronting  generalized hidden local symmetry chiral model
with the ALEPH data on the decay $\tau^-\to\pi^+\pi^-\pi^-\nu_\tau$.}
\author{N.~N.~Achasov}
\email[]{achasov@math.nsc.ru} \affiliation{Laboratory of
Theoretical Physics, S.~L.~Sobolev Institute for Mathematics,
630090, Novosibirsk, Russian Federation
}%
\author{A.~A.~Kozhevnikov}
\email[]{kozhev@math.nsc.ru} \affiliation{Laboratory of
Theoretical Physics, S.~L.~Sobolev Institute for Mathematics, and
Novosibirsk State University, 630090, Novosibirsk, Russian
Federation}

\date{\today}
\begin{abstract}
Generalized Hidden Local Symmetry (GHLS) model is the chiral model
of pseudoscalar, vector, and axial vector mesons and their interactions. It contains also
the couplings of strongly interacting particles with electroweak
gauge bosons. Here,  GHLS model  is confronted with
the ALEPH data on the decay $\tau^-\to\pi^-\pi^-\pi^+\nu_\tau$. It
is shown that the invariant mass spectrum of final pions in this decay calculated
in GHLS framework with the single $a_1(1260)$ resonance disagrees with
the experimental data at any reasonable number of free GHLS parameters.
Two modifications of GHLS model based on inclusion of two additional heavier axial vector mesons are studied.
One of them giving a good description of the ALEPH data,  with all the parameters kept free
is shown to result in very large $\Gamma_{a_1^\pm\to\pi^\pm\gamma}$ partial width. The other scheme
with  the GHLS parameters fixed in a way that the universality is preserved and the observed central value of
$\Gamma_{a_1^\pm\to\pi^\pm\gamma}$ is reached,
results in a good description of the three pion spectrum in $\tau^-\to\pi^+\pi^-\pi^-\nu_\tau$
decay.

\end{abstract}
\pacs{11.30.Rd;12.39.Fe;13.30.Eg}

\maketitle

\section{Introduction}
\label{intro}~

There is popular chiral model of   pseudoscalar, vector, and axial
vector mesons and their interactions based on nonlinear
realization of chiral symmetry, the so called Generalized Hidden
Local Symmetry (GHLS) model \cite{bando85,bando88a,bando88,meissner88}. One
of its virtue is that the sector of electroweak interactions is
introduced in such a way that the low energy relations in the
sector of strong interactions are not violated upon inclusion of
photons and electroweak gauge bosons \cite{schechter}. Some interesting two- and
three-particle decays as, for example, $\rho^0\to\pi^+\pi^-$ and
$\omega\to\pi^+\pi^-\pi^0$, were  analyzed in  the framework of
GHLS \cite{bando88}.

Some time ago GHLS  with particular choice of the renormalized [see Eq.~(\ref{renorm})]
free parameters
\begin{equation}
(a,b,c,d,\alpha_4,\alpha_5,\alpha_6)=
(2,2,2,0,-1,1,1),\label{canon}\end{equation}and $\alpha_1=\alpha_2=\alpha_3=0$, see
Refs.~\cite{bando88a,ach05,ach08a,ach08b} and (\ref{lghls}) for
more detail, was applied to the evaluation of the four-pion
process $\rho\to4\pi$ \cite{ach05,ach08a,ach08b} and to the
comparison with existing data on the reaction
$e^+e^-\to\pi^+\pi^-\pi^+\pi^-$ \cite{cmd2,babar}. It was shown
that while the results of calculations do not contradict the data
\cite{cmd2} at energies near $m_\rho$, at higher energies near 1
GeV the cross section of above reaction measured in independent
experiments \cite{cmd2,babar}, by the factor of about 30 exceeds
the values evaluated in GHLS \cite{ach08a,ach08b}. The
contributions of higher resonances $\rho^\prime$,
$\rho^{\prime\prime}$ were included to reconcile the data with
calculations \cite{ach08a,ach08b}.

Since axial vector meson $a_1(1260)$ appears only in the
intermediate states of the reaction
$e^+e^-\to\pi^+\pi^-\pi^+\pi^-$, it would be desirable to study
the processes where it manifests directly as in  the decay
$\tau^-\to\pi^+\pi^-\pi^-\nu_\tau$.  This decay  was  studied by
ALEPH Collaboration \cite{aleph05}.  The GHLS model includes  a number of
free parameters. See Section \ref{outline}. Some particular choices, as,
for example, Eq.~(\ref{canon}), were adopted in the literature.  The aim of the present paper
is to evaluate the $\pi^+\pi^-\pi^-$ spectrum in the decay of
$\tau^-$ lepton in the framework of GHLS and compare the results
with the ALEPH data. We stay with the minimal set of free parameters Eq.~(\ref{canon}) of Refs.~\cite{bando88a,bando88}.
However, contrary to the cited works, we try  to determine them from the data
and compare them with the "canonical" values Eq.~(\ref{canon}).

There are alternative attempts to apply chiral models other than GHLS one, to describe
the spectrum of three pions in $\tau$ decay. See, for example,
Refs.~\cite{gomez04,gomez09}. Application for the same purpose of purely phenomenological effective lagrangian
which does not possess the property of chiral invariance is considered in Ref.~\cite{lichard}.

The material is organized as follows. The terms  of GHLS
lagrangian necessary for calculation of the $W^-\to\pi^+\pi^-\pi^-$
decay amplitude are given in Section \ref{outline}.
Sec.~\ref{wampl} and \ref{spectrum} contain, respectively,  the expressions for the amplitude $W^-\to\pi^+\pi^-\pi^-$
and the spectrum of the state  $\pi^+\pi^-\pi^-$ side by side with the  necessary spectral functions.
Sec.~\ref{results} contains the results of
calculations of the spectrum of $\pi^+\pi^-\pi^-$ state under various
assumptions about contributions of the intermediate  axial vector mesons.
The results of evaluation of the width of the radiative  decay $a_1^\pm\to\pi^\pm\gamma$
are presented in the same section.
The discussion of the obtained results and conclusion can be found  in Sec.~\ref{concl}.

\section{Outline  of GHLS chiral lagrangian}\label{outline}
~

The basis of the derivation is the lagrangian of (GHLS)
\cite{bando88a,bando88} which includes pseudoscalar, vector, and
axial vector fields $\xi$, $V_\mu$, and $A_\mu$, respectively.
In the gauge $\xi_M=1$,
$\xi^\dagger_L=\xi_R=\xi$ and after rotating away the axial
vector-$\pi$ mixing by choosing
\begin{equation}
A_\mu=a_\mu-\frac{b_0c_0}{g(b_0+c_0)}A_{(\xi)\mu},\label{physa1}\end{equation}
where $a_\mu$ is $a_1$ meson field, $g$ is the coupling constant
to be related to $g_{\rho\pi\pi}$, and
\begin{equation}
A_{(\xi)\mu}=\frac{\partial_\mu\xi^\dag\xi-\partial_\mu\xi\xi^\dag}{2i},\label{Axi}\end{equation}
the relevant terms corresponding to  strong interactions look like
%\begin{widetext}
\begin{eqnarray}
{\cal L}_{\rm strong}&=&a_0f^{(0)2}_\pi{\rm
Tr}\left(\frac{\partial_\mu\xi^\dagger\xi+\partial_\mu\xi\xi^\dagger}{2i}-gV_\mu\right)^2+
\nonumber\\&&f^{(0)2}_\pi\left(d_0+\frac{b_0c_0}{b_0+c_0}\right){\rm
Tr}A^2_{(\xi)\mu}+\nonumber\\&&(b_0+c_0)f^{(0)2}_\pi g^2{\rm
Tr}a^2_\mu+d_0f^{(0)2}_\pi{\rm Tr}A^2_{(\xi)\mu}
-\nonumber\\&&\frac{1}{2}{\rm
Tr}\left(F^{(V)2}_{\mu\nu}+F^{(A)2}_{\mu\nu}\right)-\nonumber\\&&i\alpha_4g{\rm
Tr}[A_\mu,A_\nu]F^{(V)}_{\mu\nu}+2i\alpha_5\times\nonumber\\&&{\rm
Tr}\left(\left[A_{(\xi)\mu},A_\nu\right]
+g[A_\mu,A_\nu]\right)F^{(V)}_{\mu\nu}.
\label{lghls}\end{eqnarray}
%\end{widetext}
The lagrangian contains a number of free parameters
$a_0,b_0,c_0,d_0,\alpha_4,\alpha_5$. The counter terms with free parameters  $\alpha_{4,5}$ are
necessary for cancelation of momentum dependence in the $\rho\pi\pi$
vertex. They are chosen in accord with Refs.~\cite{bando88a,bando88} in
such a way that among the terms with higher derivatives those with
$\alpha_1,\alpha_2,\alpha_3$ are set to zero, and
only the $\alpha_{4,5,6}$ terms are included, with the
additional assumption $\alpha_5=\alpha_6$ about the arbitrary constants
multiplying the lagrangian terms. The remaining ones $\alpha_4$ and $\alpha_5$
should be related like
\begin{equation}
\alpha_4=1-\frac{2\alpha_5c_0}{b_0},\label{alpha45}\end{equation}
in order to provide the desired  cancelation.
The notations, assuming the restriction to the sector of the
non-strange mesons,  are
\begin{eqnarray}
F^{(V)}_{\mu\nu}&=&\partial_\mu V_\nu-\partial_\nu
V_\mu-ig[V_\mu,V_\nu]-ig[A_\mu,A_\nu],\nonumber\\
F^{(A)}_{\mu\nu}&=&\partial_\mu A_\nu-\partial_\nu
A_\mu-ig[V_\mu,A_\nu]-ig[A_\mu,V_\nu],\nonumber\\
V_\mu&=&\left(\frac{{\bm\tau}}{2}\cdot{\bm\rho}_\mu\right),\nonumber\\
A_\mu&=&\left(\frac{{\bm\tau}}{2}\cdot{\bm A }_\mu\right),\nonumber\\
\xi&=&\exp i\frac{{\bm\tau}\cdot{\bm\pi}}{2f^{(0)}_\pi},\label{not}\end{eqnarray} where
${\bm\rho}_\mu$, ${\bm\pi}$ are the vector meson $\rho$  and
pseudoscalar pion fields, respectively, ${\bm A}_\mu$ is the axial
vector field [not $a_1$ meson, see Eq.~(\ref{physa1})],
${\bm\tau}$ is the isospin Pauli matrices. Free parameters
$(a_0,b_0,c_0,d_0)$, and $f^{(0)}_\pi$ of the GHLS lagrangian with
index $0$ are bare parameters before renormalization (see below);
$[,]$ stands for commutator. Hereafter the boldface characters,
cross ($\times$), and dot ($\cdot$) stand for vectors, vector
product, and scalar product, respectively, in the isotopic space.

One should notice that in distinction with
Refs.~\cite{bando88a,bando88} where only the linear piece $A_{(\xi)\mu}\propto{\bm\tau}\cdot\partial_\mu{\bm\pi}$,
is rotated away, we, first, rotate away the nonlinear combination
Eq.~(\ref{Axi}). As was shown earlier \cite{ach05}, it results in the amplitude of the decay
$a_1\to3\pi$ satisfying the Adler condition even for the off-mass-shell
$a_1$ meson. Second,  at no point we use the equations of motion of free
fields. This is because the axial, vector, and pseudoscalar mesons
are often outside their respective mass shells in the process considered in the
present paper.

GHLS lagrangian includes also electroweak sector. In what follows
we will neglect the terms quadratic in electroweak coupling
constants keeping only the terms linear in above couplings. These
terms describe the interaction of $\pi$, $\rho$, and $a_1$ mesons
with electroweak gauge bosons and look as \cite{bando88a,bando88}
%\begin{widetext}
\begin{eqnarray}
\Delta{\cal L}_{\rm EW}&=&2f_\pi^{(0)2}\bar{g}{\rm
Tr}\left\{a_0\left(\frac{\partial_\mu\xi^\dag\xi+\partial_\mu\xi\xi^\dag}{2i}\times\right.\right.
\nonumber\\&&\left.\left.\frac{\xi^\dag{\cal L}_\mu\xi+\xi{\cal
R}_\mu\xi^\dag}{2}\right)+\left(d_0+\frac{b_0c_0}{b_0+c_0}\right)\times\right.\nonumber\\&&\left.
A_{(\xi)\mu}\frac{\xi^\dag{\cal L}_\mu\xi-\xi{\cal
R}_\mu\xi^\dag}{2}-\right.\nonumber\\&&\left.a_0gV_\mu\frac{\xi^\dag{\cal
L}_\mu\xi+\xi{\cal
R}_\mu\xi^\dag}{2}+\right.\nonumber\\&&\left.b_0ga_\mu\frac{\xi^\dag{\cal
L}_\mu\xi-\xi{\cal R}_\mu\xi^\dag}{2}\right\}.\label{LEW}
\end{eqnarray}
%\end{widetext}
Here we keep only the charged electroweak sector, hence
\cite{bando88a,bando88},
\begin{eqnarray}
\bar{g}{\cal L}_\mu&=&\frac{g_2}{\sqrt{2}}(W^+_\mu
T_-+W^-_\mu T_+),\nonumber\\
\bar{g}{\cal R}_\mu&=&0, \label{LR}\end{eqnarray}  $W^\pm_\mu$ are the
fields of $W^\pm$ bosons,  $g_2$ is the  electroweak $SU(2)$ gauge
coupling constant. In the $SU(2)$ subgroup of the flavor $SU(3)$
group of strong interactions,
\begin{equation}
T^+=\left(%
\begin{array}{cc}
  0 & V_{ud} \\
  0 & 0 \\
\end{array}%
\right),\label{T+}\end{equation} $V_{ud}=\cos\theta_C$ is the
element of Cabibbo-Kobayashi-Maskawa matrix.

In the spirit of chiral perturbation theory, as the first step in
obtaining necessary terms, one should expand the matrix $\xi$ into
the series over ${\bm\pi}/f_\pi^{(0)}$. The second step is the
renormalization necessary for canonical normalization of the pion
kinetic term. The renormalization is \cite{bando88a,bando88}
\begin{eqnarray}
f^{(0)}_\pi&=&Z^{-1/2}f_\pi, {\bm\pi}\to Z^{-1/2}{\bm\pi},
(a_0,b_0,c_0,d_0)=\nonumber\\&&Z\times(a,b,c,d),\label{renorm}\end{eqnarray} where
$$\left(d_0+\frac{b_0c_0}{b_0+c_0}\right)Z^{-1}=1.$$

Close examination of Eq.~(\ref{LEW}) shows that the expansion
includes the point-like interaction
$$\left(\frac{a}{2}-d-\frac{bc}{b+c}\right)W^-_\mu
[{\bm\pi}\times\partial_\mu{\bm\pi}]_{1+i2}.$$ Analogous term
appears when one restores electromagnetic field. Since there are
no experimental indications on point-like $\gamma\to\pi^+\pi^-$
vertex, we set
\begin{equation}
\frac{a}{2}-d-\frac{bc}{b+c}=0.\label{nopoint}\end{equation} This
relation removes also the above point-like $W^-\to\pi^-\pi^0$
vertex.

\section{The amplitude of the transition $W^-\to2\pi^-\pi^+$.}
\label{wampl} ~

As for the  strong interaction sector, part of necessary  terms of
the low momentum expansion concerning the transition $a_1\to3\pi$
which are  relevant for the present work were given in
Ref.~\cite{ach05}, assuming the "canonical" choice of free
parameters (\ref{canon}). Let us rewrite them without such
assumption. First note that
\begin{equation}
g_{\rho\pi\pi}=\frac{ag}{2},\label{grhopp}\end{equation}
\begin{equation}
m^2_\rho=ag^2f^2_\pi,\label{mrho}\end{equation}
\begin{equation}
m^2_{a_1}=(b+c)g^2f^2_\pi,\label{ma1}\end{equation}where
$f_\pi=92.4$ MeV is the pion decay constant. Notice that we fix hereafter
$g_{\rho\pi\pi}$ from the experimental value of the $\rho^0\to\pi^+\pi^-$ decay width leaving  $a$
as free parameter. Second, the
lagrangian describing the decay $a_1\to3\pi$ can be written as
\begin{eqnarray}
{\cal L}_{a_13\pi}&=&-\frac{r}{f_\pi}(\partial_\mu{\bm
a}_\nu-\partial_\nu{\bm
a}_\mu)[{\bm\rho}_\mu\times\partial_\nu{\bm\pi}]+\nonumber\\&&
\frac{\alpha_5}{f_\pi}{\bm
a}_\mu[(\partial_\mu{\bm\rho}_\nu-\partial_\nu{\bm\rho}_\mu)\times\partial_\nu{\bm\pi}]-\nonumber\\&&
\frac{r^2}{gf^3_\pi}(\alpha_5-r)[{\bm
a}_\mu\times\partial_\nu{\bm\pi}]\cdot[\partial_\mu{\bm\pi}\times\partial_\nu{\bm\pi}]-\nonumber\\&&
\frac{r}{2gf^3_\pi}\partial_\mu{\bm
a}_\nu\cdot[{\bm\pi}\times[\partial_\mu{\bm\pi}\times\partial_\nu{\bm\pi}]].\label{La13pi}
\end{eqnarray}
The amplitude of the decay
$a_1^-(q)\to\pi^+(q_1)\pi^-(q_2)\pi^-(q_3)$ calculated from
Eq.~(\ref{La13pi}) can be written as follows:
$M[a_1^-(q)\to\pi^+(q_1)\pi^-(q_2)\pi^-(q_3)]\equiv M_{a_13\pi}$,
\begin{eqnarray}
iM_{a_13\pi}&=&\frac{agr}{2f_\pi}
\epsilon_\mu\left(A_1q_{1\mu}+A_2q_{2\mu}+A_3q_{3\mu}\right),
\label{ma13pi}\end{eqnarray} where $\epsilon_\mu$ is the
polarization four-vector of $a_1$ meson, and
\begin{widetext}
\begin{eqnarray}
A_1&=&(1+\hat{P}_{23})\left\{\frac{\beta[(q_3,q_1-q_2)-(q,q_3)+m^2_\pi]-(q,q_3)}{D_\rho(q_1+q_2)}
+\frac{4r^2(\beta-1)(q_2,q_3)+(q,q)-(q,q_1)}{2m^2_\rho}\right\},\nonumber\\
A_2&=&\frac{\beta[(q_3,q_1-q_2)+(q,q_3)-m^2_\pi]+(q,q_3)}{D_\rho(q_1+q_2)}+\frac{(q_2,q_1-q_3)}{D_\rho(q_1+q_3)}
-\frac{2r^2(\beta-1)(q_1,q_3)+(q,q_1)}{m^2_\rho}. \label{A12}
\end{eqnarray}
\end{widetext}
Hereafter $\hat{P}_{ij}$ interchanges pion momenta $q_i$ and
$q_j$, $(q_i,q_j)$ stands for the Lorentz scalar product of  four-vectors, and
$A_3=\hat{P}_{23}A_2$. Parameters $r$ and $\beta$ are the
combinations of the GHLS parameters:
\begin{eqnarray}
r&=&\frac{b}{b+c}\mbox{,
}\beta=\frac{\alpha_5}{r}.\label{rbet}\end{eqnarray}
Notice that the amplitude (\ref{ma13pi}) respects the Adler
condition \cite{adler65}: it vanishes in the chiral limit $m^2_\pi\to 0$
when the four-momentum of any final pion  vanishes \cite{ach05}. Such a property
is the manifestation of the chiral invariance.

\begin{figure}
\includegraphics[width=80mm]{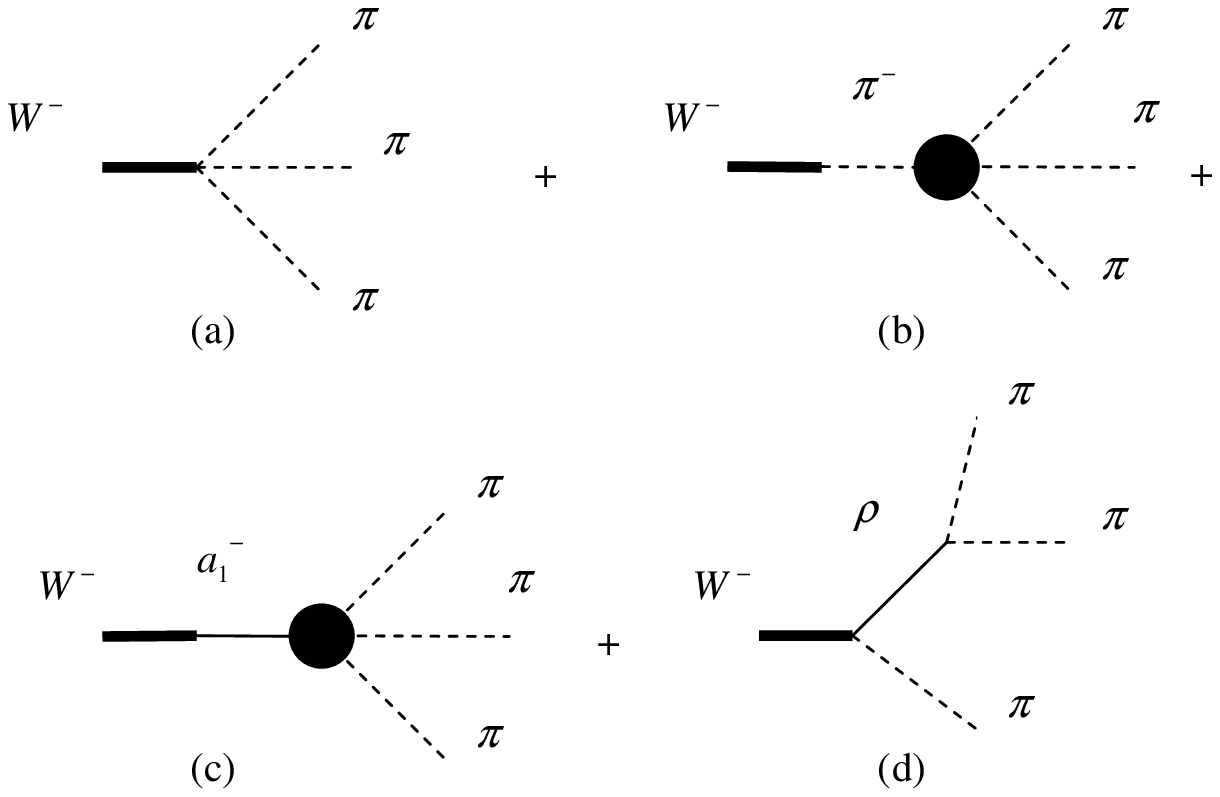}
\caption{\label{taucdiag}Diagrams schematically describing the transition
$W^-\to\pi^-\pi^-\pi^+$. Shaded circles depict the transition
including both  the point-like and $\rho$-exchange contributions.
Permutations of pion momenta are understood.}
\end{figure}

The amplitude of the decay $\tau^-\to\pi^-\pi^-\pi^+\nu_\tau$
incorporates the transition $W^-\to\pi^-\pi^-\pi^+$. In GHLS, the
latter is given by the diagrams shown in Fig.~\ref{taucdiag}.
Necessary terms are obtained from the low momentum expansion of
electroweak piece of GHLS lagrangian Eq.~(\ref{LEW}) and look like
\begin{eqnarray}
\Delta{\cal L}_{\rm EW}&=&\frac{1}{2}g_2V_{ud}{\bm
W}_{\mu\bot}\left(-f_\pi\partial_\mu{\bm\pi}_\bot+\right.\nonumber\\&&\left.\frac{1}{3f_\pi}
[{\bm\pi}\times[{\bm\pi}\times\partial_\mu{\bm\pi}]]_\bot
+\right.\nonumber\\&&\left.bgf^2_\pi{\bm a}_{\mu\bot}
+agf_\pi[{\bm\pi}\times{\bm\rho}_\mu]_\bot\right), \label{DeltaL}
\end{eqnarray}
where  the vector ${\bm V}_\bot=(V_1,V_2)$ denotes transverse
charged components of the isotopic vector.
%%%%%%%%%%%%%%%%%%%%%%%%%%%%%%%%%%%%%%%%%%%%%%%%%%%%%%%%%%%%%%%%%%%
%%%%%%%%%%%%%%%%%%%%%%%%%%%%%%%%%%%%%%%%%%%%%%%%%%%%%%%%%%%%%%%%%%%%
The amplitude of the decay
$W^-(q)\to\pi^+(q_1)\pi^-(q_2)\pi^-(q_3)$ corresponding to the
diagrams Fig.~\ref{taucdiag} is
\begin{equation}
iM=\frac{g_2V_{ud}}{2f_\pi}\epsilon_\mu^{(W)}J_\mu,\label{M}\end{equation}
where $\epsilon_\mu^{(W)}$ is the polarization four-vector of
$W^-$ boson and the axial decay current $J_\mu$ looks like
\begin{eqnarray}
J_\mu&=&-q_{1\mu}+\frac{q_\mu}{D_\pi(q)}\left[m^2_\pi-(q,q_1)+\frac{am^2_\rho}{2}\times\right.\nonumber\\&&
\left.(1+\hat{P}_{23})\frac{(q_2,q_1-q_3)}{D_\rho(q_1+q_3)}\right]-\frac{ar^2m^2_{a_1}}{2D_{a_1}(q)}\times\nonumber\\&&
\left\{A_1q_{1\mu}+A_2q_{2\mu}+A_3q_{3\mu}-\frac{2q_\mu}{m^2_{a_1}}\times\right.\nonumber\\&&\left.(1+\hat{P}_{23})
\left[(m^2_\pi+(q_1,q_2))(q_3,q_1-q_2)\times\right.\right.\nonumber\\&&\left.\left.
\left(\frac{\beta}{D_\rho(q_1+q_2)}-\frac{r^2(\beta-1)}{m^2_\rho}\right)\right]\right\}+\nonumber\\&&\frac{am^2_\rho}{2}
(1+\hat{P}_{23})\frac{(q_1-q_3)_\mu}{D_\rho(q_1+q_3)}.
\label{Jdec}
\end{eqnarray}
In the above expressions, $D_\rho$, $D_\pi$, and $D_{a_1}$ are the
inverse propagators of $\pi$, $\rho$, and $a_1$ mesons,
respectively. Their expressions are given in Ref.~\cite{ach05}.
The terms corresponding to the diagrams (a), (b), (c), and (d) in
Fig.~\ref{taucdiag} are easily identified by these propagators.

\section{The spectrum of $\pi^+\pi^-\pi^-$ in $\tau^-\to\pi^+\pi^-\pi^-\nu_\tau$
decay}\label{spectrum}
~

The spectrum of the three pion state in the decay
$\tau^-\to\pi^+\pi^-\pi^-\nu_\tau$ normalized to its branching
fraction is \cite{okun}
\begin{eqnarray}
\frac{dB}{ds}&=&\frac{(G_FV_{ud})^2(m^2_\tau-s)^2}{2\pi(2m_\tau)^3\Gamma_\tau}
\times\nonumber\\&&\left[(m^2_\tau+2s)\rho_t(s)+m^2_\tau\rho_l(s)\right],\end{eqnarray}
$s=q^2$, $G_F$ is the Fermi constant, and $\Gamma_\tau$ is the
width of $\tau$ lepton. The transverse and longitudinal spectral
functions are, respectively,
\begin{eqnarray}
\rho_t(s)&=&\frac{1}{3\pi sf^2_\pi}\int
d\Phi_{3\pi}\left[\frac{|(q,J)|^2}{s}-(J,J^\ast)\right],\nonumber\\
\rho_l(s)&=&\frac{1}{\pi s^2f^2_\pi}\int d\Phi_{3\pi}|(q,J)|^2,
\end{eqnarray}
where $d\Phi_{3\pi}$ is the element of Lorentz-invariant phase
space  volume of the system $\pi^-\pi^-\pi^+$.
%%%%%%%%%%%%%%%%%%%%%%%%%%%%%%%%%%%%%%%%%%%%%%%%%%%%%%%%%%%%%%%%%%%
\begin{figure}
\includegraphics[width=75mm]{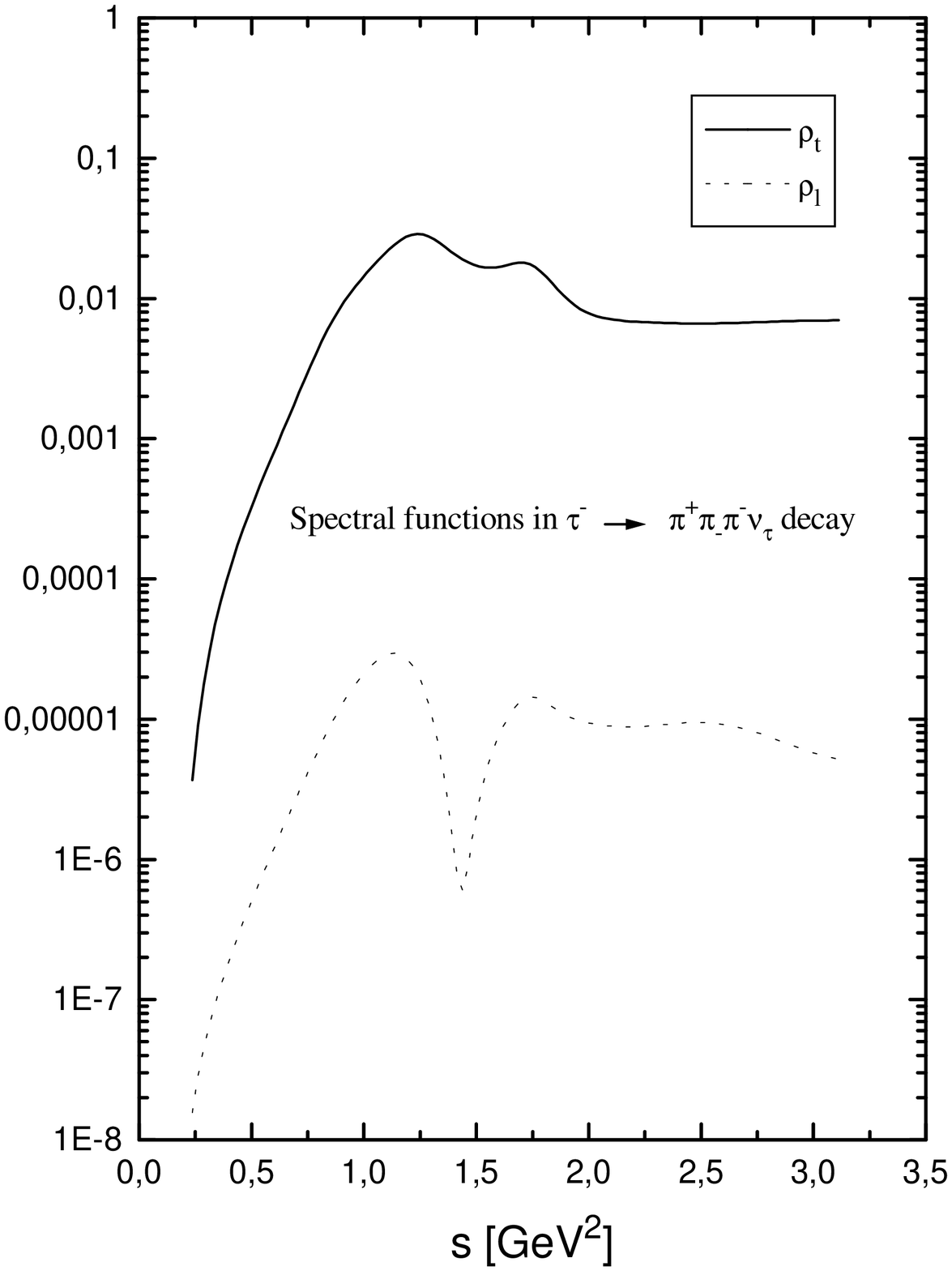}
\caption{\label{rhotl}The transverse $\rho_t$ and longitudinal
$\rho_l$ spectral functions in  $\tau^-\to\pi^+\pi^-\pi^-\nu_\tau$
decay evaluated in GHLS model with the single axial vector meson
under assumption of "canonical" choice (\ref{canon}) of free
parameters.  See the text for more
detail.}
\end{figure}
%%%%%%%%%%%%%%%%%%%%%%%%%%%%%%%%%%%%%%%%%%%%%%%%%%%%%%%%%%%%%%%%%%%%
The numerical integration shows that
$\rho_l$ is by about three orders of magnitude smaller than
$\rho_t$ in all allowed kinematical range $9m^2_\pi<s<m^2_\tau$.
See Fig.~\ref{rhotl}. By  this reason it is neglected in what
follows.

\section{Results}
\label{results} ~

\begin{figure}
\includegraphics[width=85mm]{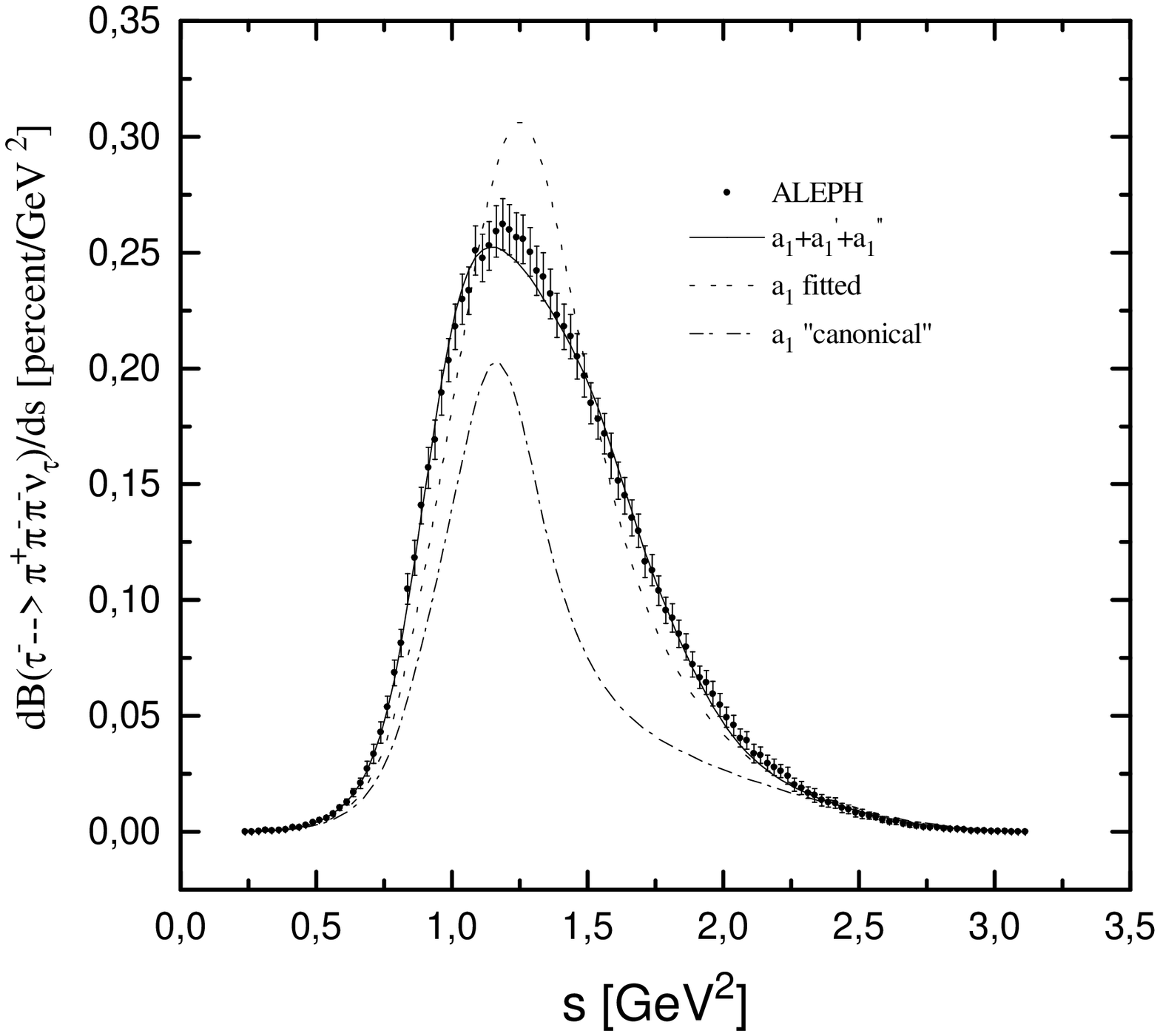}
\caption{\label{tau3pic}The spectrum of the decay  $\pi^+\pi^-\pi^-$ state in $\tau$
normalized to the branching fraction
$B_{\tau^-\to\pi^-\pi^-\pi^+\nu_\tau}$. The solid line is drawn
for the set of GHLS parameters of the variant A of the Table
\ref{fitresults}. The ALEPH data are from
Ref.~\cite{aleph05}. See the text for more  detail.}
\end{figure}

The "canonical" choice (\ref{canon})  of free GHLS parameters
\cite{bando88} with $m_{a_1}=1.23$ GeV results in the spectrum
shown with the dot-dashed line in Fig.~\ref{tau3pic}.
It disagrees with the data both in lower branching ratio  $B_{\tau^-\to\pi^+\pi^-\pi^-\nu_\tau}
\approx6\%$ and in the shape of the spectrum.  Upon
the variation of  free parameters of the single $a_1$ resonance
contribution listed in Eq.~(\ref{canon})  one obtains  the curve drawn in Fig.~\ref{tau3pic}
with the dashed  line. Corresponding parameters $m_{a_1}\approx1.54\mbox{ GeV, }a\approx1.75\mbox{,
}r\approx1.05\mbox{, }\beta\approx0.84$ with $\chi^2/N_{\rm d.o.f.}=690/112$.
reproduce the branching ratio $B_{\tau^-\to\pi^+\pi^-\pi^-\nu_\tau}
\approx9\%$ but the shape of the spectrum is not reproduced.
Inclusion of additional higher derivative terms to the suggested in
Refs.~\cite{bando88a,bando88} minimal set Eq.~(\ref{canon})  and subjected to the fitting in the present work
cannot improve the situation. Indeed, even the minimal set Eq.~(\ref{canon})
results in a rather fast growth of the $a_1\to3\pi$ decay width with the energy
increase, see Ref.~\cite{ach05} and Fig.~\ref{widthsa1} and \ref{gaa1fr} below in Sec.~\ref{concl}.
Additional higher derivative terms would make
the growth to be explosive. Restricting such a growth would require phenomenological
form factors with free parameters. We believe that the dynamical explanation
of the shape of the spectrum  based on additional axial vector resonances $a_1^\prime$,
$a_1^{\prime\prime}$ would be preferable. Note that there are indications on such resonances,
both theoretical \cite{isgur87,isgur85} and experimental \cite{pdg,amelin,cleo}.

Hence, to improve the fit, we include the contributions of heavier axial
vector  resonances $a_1^\prime$,
$a_1^{\prime\prime}$.  Taking them into account  reduces to adding two diagrams
similar to one in Fig.~\ref{taucdiag}(c), with the replacement of $a_1(1260)$
by $a^\prime_1$ and $a^{\prime\prime}_1$. Since there is
no available information concerning their couplings,
the above resonances  are  included in a way analogous to $a_1(1260)$.
%%%%%%%%%%%%%%%%%%%%%%%%%%%%%%%%%%%%%%%%%%%%%%%%%%%%%%%%%%%%%%%%%%%
%%%%%%%%%%%%%%%%%%%%%%%%%%%%%%%%%%%%%%%%%%%%%%%%%%%%%%%%%%%%%%%%%%%%
This prescription results in the amplitudes of the decays $a^\prime_1,a^{\prime\prime}_1\to3\pi$
vanishing when the four-momentum of any final pion vanishes. That is,
the Adler condition is not violated upon adding the above resonances. In this sense
the way of inclusion them respects chiral symmetry.
%%%%%%%%%%%%%%%%%%%%%%%%%%%%%%%%%%%%%%%%%%%%%%%%%%%%%%%%%%%%%%%%%%%
\begin{figure}
\includegraphics[width=85mm]{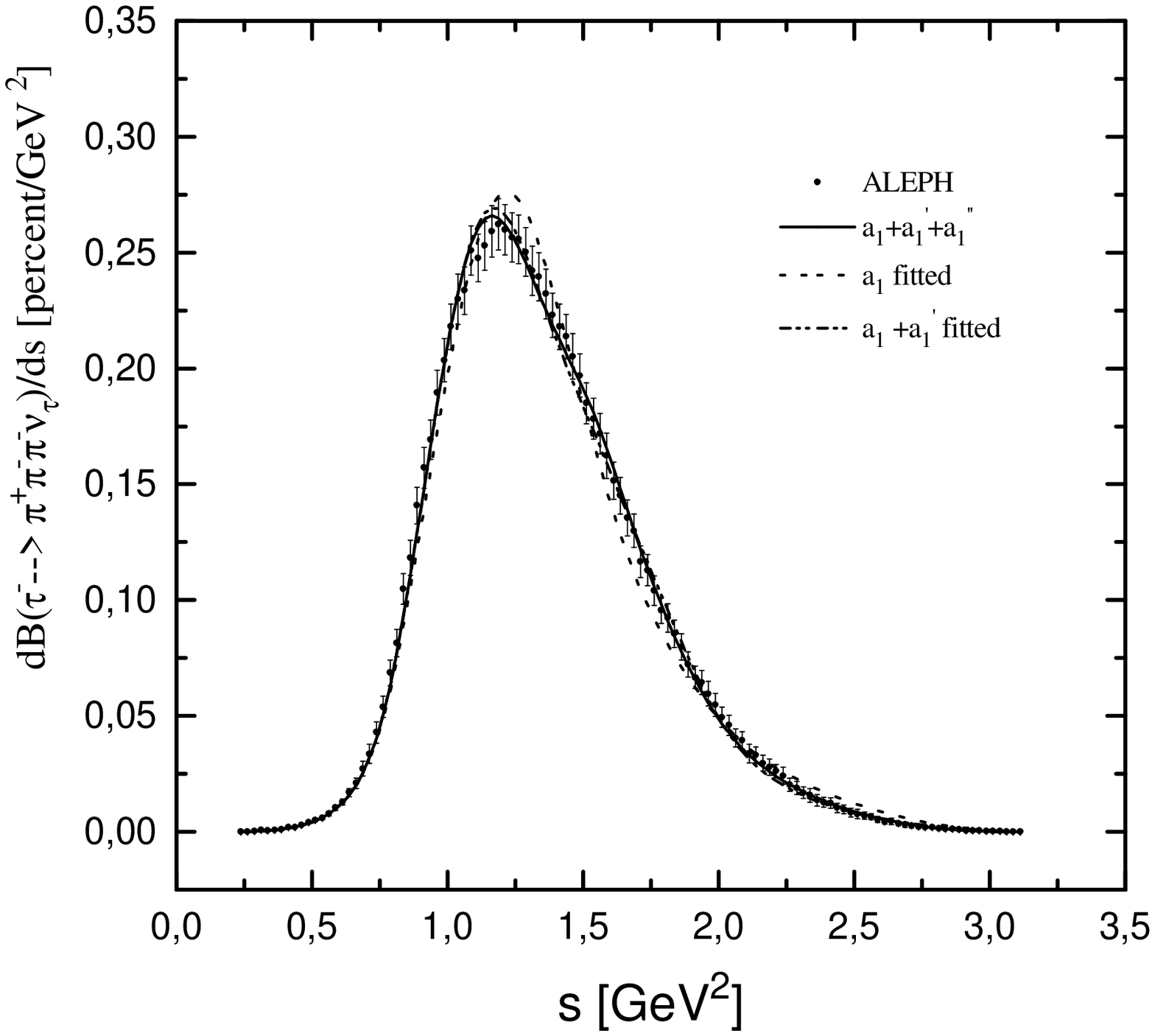}
\caption{\label{fixr}The spectrum of the decay $\pi^+\pi^-\pi^-$ in $\tau$
normalized to the branching fraction
$B_{\tau^-\to\pi^-\pi^-\pi^+\nu_\tau}$. It corresponds to the
variants C and D in the Table \ref{fitresCD}.
The ALEPH data are from
Ref.~\cite{aleph05}. See the text for more  detail.}
\end{figure}
%%%%%%%%%%%%%%%%%%%%%%%%%%%%%%%%%%%%%%%%%%%%%%%%%%%%%%%%%%%%%%%%%%%%

The total set of the fitted parameters is
first taken to be
$$(m_{a_1},a,r,\beta,m_{a^\prime_1},a^\prime,r^\prime,\beta^\prime,w^\prime,
m_{a^{\prime\prime}_1},a^{\prime\prime},r^{\prime\prime},\beta^{\prime\prime},w^{\prime\prime}).$$
The parameters $a^\prime$, $r^\prime$, $\beta^\prime$ characterize the $a^\prime_1\to3\pi$
decay amplitude similar to Eq.~(\ref{ma13pi}), (\ref{A12}) in the case of $a_1(1260)\to3\pi$, while
$w^\prime$  parameterizes  the coupling $a^\prime_1\rho\pi$
as $g_{\rho\pi\pi}w^\prime r^\prime/f_\pi$. Compare with
Eq.~(\ref{ma13pi}). Analogously for $a_1^{\prime\prime}$.  The fit
chooses $w^\prime=1$ and turns out to be insensitive to this
parameter leaving $\chi^2/N_{\rm d.o.f}=122/102$. The quality of the fit
can be considerably improved upon fixing $w^\prime=1$ but adding new
parameter $\psi^\prime$-the phase of the $a^\prime_1$
contribution. Such phase imitates possible mixing among $a_1$,
$a_1^\prime$, $a_1^{\prime\prime}$ resonances. The results of such
type of the fit are given in the column variant A of the Table \ref{fitresults}.
Corresponding curve is shown in
Fig.~\ref{tau3pic} with the solid line.
Using Eqs.~(\ref{grhopp}), (\ref{mrho}), (\ref{ma1}), (\ref{rbet}), and obtaining
$g_{\rho\pi\pi}=5.95$ from $\Gamma_{\rho\pi\pi}$ \cite{pdg} one can
compare the fitted GHLS parameters with the "canonical" ones
Eq.~(\ref{canon}). To this end one should invoke the condition of
cancelation Eq.~(\ref{nopoint}) of the point-like
$\gamma\pi^+\pi^-$ and $W^-\pi^-\pi^0$ vertices in GHLS.
The relations expressing  the original GHLS parameters through the fitted ones are
the following:
\begin{eqnarray}
b&=&r\left(\frac{m_{a_1}a}{2f_\pi g_{\rho\pi\pi}}\right)^2,\nonumber\\
c&=&(1-r)\left(\frac{m_{a_1}a}{2f_\pi g_{\rho\pi\pi}}\right)^2,\nonumber\\
d&=&\frac{a}{2}-r(1-r)\left(\frac{m_{a_1}a}{2f_\pi g_{\rho\pi\pi}}\right)^2,\nonumber\\
\alpha_4&=&1-2\beta(1-r),\nonumber\\
\alpha_5&=&\beta r,\nonumber\\
\alpha_6&=&\alpha_5.
\label{realparam}\end{eqnarray}
These GHLS parameters are marked in the Table \ref{fitresults} as "calculated".
Since the basis of inclusion of  heavier resonances $a_1^\prime$ and $a_1^{\prime\prime}$
here is  purely phenomenological, specifically, there is no analog of gauge coupling constant
$g$, we do not recalculate $(a^\prime,b^\prime,c^\prime,d^\prime,\cdots)$ and
$(a^{\prime\prime},b^{\prime\prime},c^{\prime\prime},d^{\prime\prime},\cdots)$ similar to
Eq.~(\ref{realparam}).
%%%%%%%%%%%%%%%%%%%%%%%%%%%%%%%%%%%%%%%%%%%%
\begin{table}%[H] add [H] placement to break table across pages
\caption{\label{fitresults}The values of free parameters of GHLS model obtained from
the unconstrained fit of the ALEPH data  on the decay $\tau^-\to\pi^+\pi^-\pi^-\nu_\tau$
\cite{aleph05} (variant A), and the fit with
the constrain $a=2$ preserving universality (variant B). Also shown are the corresponding
calculated original \cite{bando88a,bando88} GHLS parameters
and the   magnitudes of branching fractions of the above decay.  }
\begin{ruledtabular}
\begin{tabular}{ccc}
parameter&variant A&variant B\\
\hline
$m_{a_1}$[GeV]&$1.332\pm0.015$&$1.139\pm0.016$\\
$a$&$1.665\pm0.011$&$\equiv 2$\\
$b$(calculated)&$1.35\pm0.05$&$0.52\pm0.03$\\
$c$(calculated)&$2.72\pm0.08$&$3.74\pm0.11$\\
$d$(calculated)&$-0.07\pm0.03$&$0.54\pm0.03$\\
$\alpha_4$(calculated)&$-10\pm1$&$-27\pm2$\\
$\alpha_5$(calculated)&$2.82\pm0.06$&$1.94\pm0.15$\\
$r$&$0.332\pm0.007$&$0.122\pm0.006$\\
$\beta$&$8.5\pm0.3$&$15.9\pm0.9$\\
$m_{a^\prime_1}$[GeV]&$1.59\pm0.01$&$1.76\pm0.01$\\
$a^\prime$&$0.99\pm0.01$&$1.09\pm0.01$\\
$r^\prime$&$0.96\pm0.01$&$0.90\pm0.01$\\
$\beta^\prime$&$0.07\pm0.02$&$0.28\pm0.02$\\
$w^\prime$&$\equiv1$&$\equiv1$\\
$\psi^\prime$&$28^\circ\pm1^\circ$&$48^\circ\pm1^\circ$\\
$m_{a^{\prime\prime}_1}$[GeV]&$1.88\pm0.02$&$2.27\pm0.02$\\
$a^{\prime\prime}$&$0.46\pm0.01$&$0.59\pm0.01$\\
$r^{\prime\prime}$&$1.45\pm0.02$&$1.56\pm0.02$\\
$\beta^{\prime\prime}$&$0.91\pm0.05$&$0.91\pm0.03$\\
$w^{\prime\prime}$&$1.14\pm0.01$&$1.37\pm0.01$\\
$\psi^{\prime\prime}$&$\equiv0^\circ$&$\equiv0^\circ$\\
$B_{\tau^-\to\pi^+\pi^-\pi^-\nu_\tau}$&$(9.05\pm0.16)\%$&$(9.00\pm0.15)\%$\\
$\chi^2/N_{\rm d.o.f}$&79/102&70/103\\
\end{tabular}
\end{ruledtabular}
\end{table}

One can see that the obtained $a=1.665\pm0.011$ is in disagreement with the universality
condition $g_{\rho\pi\pi}=g$, which demands $a=2$, see Eq.~(\ref{grhopp}).
Hence we  fulfill also the partially constrained fit with $a\equiv2$, in order to preserve universality
of the $\rho$ couplings. The results are presented as the variant B in the Table \ref{fitresults}.
Since the shape of the spectrum in the variant B is the same as in the Fig.~\ref{tau3pic}, we do not show
corresponding curve here.

Of a special interest is the width of the radiative decay $a_1^\pm\to\pi^\pm\gamma$.
In order to evaluate the amplitude of this decay one should take into account the electromagnetic
interaction.  Upon neglecting the weak neutral current contribution, this is reduced to adding the terms
$$\bar g{\cal L}_\mu^{\rm e.m.}=eQ{\cal A}_\mu,$$ $$\bar g{\cal R}_\mu^{\rm e.m.}=eQ{\cal A}_\mu$$
to the right hand side of the first and second lines of Eq.~(\ref{LR}), respectively.
Here, ${\cal A}_\mu$ stands for the field of the photon,  $e$ is the elementary charge, and
$$Q=\frac{1}{3}\left(%
\begin{array}{cc}
  2 & 0 \\
  0 & -1 \\
\end{array}%
\right)$$ is the charge matrix restricted to the sector of nonstrange mesons.
As is known \cite{bando88a,bando88},
the above decay originates from two sources. First, the $a_1\to\rho\pi$
transition followed by the transition $\rho\to\gamma$ which is given, upon
neglecting the corrections to the masses of the second order
in the electric charge, by the $\gamma\rho^0$ mixing term
\begin{equation}
{\cal L}_{\gamma\rho}=-eagf^2_\pi\rho^0_\mu{\cal A}_\mu.\label{gammarho}\end{equation}
Second, one should add
the direct $a_1\to\pi\gamma$ transition given by the term
\begin{equation}
{\cal L}_{a_1\pi\gamma}=-iebf_\pi{\cal A}_\mu(a^+_{1\mu}\pi^--a^-_{1\mu}\pi^+).\label{a1pigamma}\end{equation}
The resulting $a_1^\pm\to\pi^\pm\gamma$ decay width is represented in the form
\begin{eqnarray}
\Gamma_{a^\pm_1\to\pi^\pm\gamma}&=&\frac{\alpha am^3_{a_1}}{24m^2_\rho}\left[r(\beta-1)\right]^2
\left(1-\frac{m^2_\pi}{m^2_{a_1}}\right)^3,\label{Gammaa1pigamma}\end{eqnarray}
where $\alpha$ is the fine structure constant. Notice, that the above expression for
$\Gamma_{a^\pm_1\to\pi^\pm\gamma}$ is written with the counter terms taken into account.
The $a^\pm_1\to\pi^\pm\gamma$ decay amplitude without counter terms is
proportional to the combination  $b-arm^2_{a_1}/m^2_\rho$ which vanishes at any choice of
GHLS parameters because of
the relations (\ref{mrho}), (\ref{ma1}), and (\ref{rbet}).
The cancelation is due to the compensation of the  diagram with the direct
transition $a_1\to\pi^\pm\gamma$ expressed by the lagrangian Eq.~(\ref{a1pigamma})
and one with the intermediate $\rho$ meson
$a_1^\pm\to\rho^0\pi^\pm$ followed by the transition $\rho^0\to\gamma$, Eq.~(\ref{gammarho}) \cite{bando88a,bando88}.

The evaluation of $\Gamma_{a^\pm_1\to\pi^\pm\gamma}$ with the parameters from the variants
A and B of the Table \ref{fitresults} gives the figures of the order of few MeV
due to large values of $\beta$  in the Table \ref{fitresults} chosen by the fits.
This is in disagreement with the  measured \cite{ziel}
\begin{equation}
\Gamma_{a^\pm_1\to\pi^\pm\gamma}=640\pm246\mbox{ keV}.\label{a1radwid}\end{equation}
Hence, one should further constrain the fit in order to incorporate the above  radiative width.
With the accuracy better than $4\%$ in $\Gamma_{a^\pm_1\to\pi^\pm\gamma}$ one can neglect
the ratio $m^2_\pi/m^2_{a_1}$ and express the parameter $\beta$ as follows:
\begin{equation}
\beta=1+\frac{m_\rho}{rm_{a_1}}\left(\frac{24\Gamma_{a^\pm_1\to\pi^\pm\gamma}}{\alpha am_{a_1}}\right)^{1/2}.
\label{beta}
\end{equation}
When fitting, the central value of Eq.~(\ref{a1radwid}) is used. In addition, $a=2$ is kept  fixed in order to
provide the universality of the $\rho$ couplings.

It is found out that the fit with the fixed  parameters $a$ and $\beta$  gives rather poor
description with $\chi^2/N_{\rm d.o.f}=209/102$. The peculiar feature of the fit is that it chooses
$\psi^\prime\approx0$, the phase of the $a_1^\prime$ contribution, but $\chi^2$ is almost
insensitive to the rather wide  variations around above central value.
Hence, we fix $\psi^\prime\equiv0$, but introduce a new free parameter $\gamma$ whose
meaning is $\gamma=m_{a_1}\Delta\Gamma_{a_1}$, where $\Delta\Gamma_{a_1}$ effectively takes into account
the contributions to the $a_1$ resonance width other than $\rho\pi+3\pi\to3\pi$ one, for example,
$a_1\to\rho^\prime\pi\to3\pi$, $K\bar K\pi$. They may be effective for the off-mass-shell $a_1$ meson. Of course,
the approximation of the constant width for these contributions is oversimplified, but it nevertheless
gives the rough estimate of their possible role as compared to the main contribution $a_1\to\rho\pi+3\pi\to3\pi$
whose energy dependence is fully taken into account.

The results of the fit  constrained by the conditions  of the universality of
the $\rho$ coupling and the fixed central value of $\Gamma_{a_1^\pm\to\pi^\pm\gamma}$
Eq.~(\ref{a1radwid}) are presented as  the variant C in the Table \ref{fitresCD}. The spectrum
of the system $\pi^+\pi^-\pi^-$ evaluated with the parameters of variant C is shown with the solid line
in Fig.~\ref{fixr}. Note that the found $\gamma=m_{a_1}\Delta\Gamma_{a_1}\sim0.3$ GeV$^2$ corresponds
to the portion of the $a_1$ decay channels different from $\rho\pi+3\pi\to3\pi$ one,
at the level $\Delta\Gamma_{a_1}/\Gamma_{a_1\to3\pi}\sim0.02$. This estimate
can be obtained from the solid curve in Fig.~\ref{widthsa1} (or Fig.~\ref{gaa1fr}), where  the
calculated $\Gamma_{a_1\to\rho\pi+3\pi\to3\pi}$ is shown. The above estimate demonstrates that
the additional contribution to the $a_1$ width beside the GHLS one
is very small.

Further evaluation shows that the contribution of
the resonance $a_1^{\prime\prime}$, see Fig.~\ref{fixr_const} and sec.~\ref{concl} below, is rather small.
Hence, we fulfill the fit in which the contribution of the
resonance is absent. The parameters found in such type of the fit are listed
in the the column variant D of the Table \ref{fitresCD}. The branching ratio
$B_{\tau^-\to\pi^+\pi^-\pi^-\nu_\tau}$ and the visual shape of the  spectrum in the variant D
are in reasonable agreement with the data, but the magnitude of $\chi^2/N_{\rm d.o.f}$ is larger by the factor of two
as compared to the variant C. This description is achieved with rather large phase of
the $a_1^\prime$ contribution $\psi^\prime\approx40^\circ$ pointing to a rather strong
$a_1a_1^\prime$ mixing in the  considered variant D.  The curve corresponding
to the variant D is shown in Fig.~\ref{fixr} with the dot-dash line.

%%%%%%%%%%%%%%%%%%%%%%%%%%%%%%%%%%%%%%%%%%%%
\begin{table}%[H] add [H] placement to break table across pages
\caption{\label{fitresCD}The values of free parameters of GHLS model obtained from
the  fit of the ALEPH data  on the decay $\tau^-\to\pi^+\pi^-\pi^-\nu_\tau$
\cite{aleph05} constrained in a way as to fix $a\equiv2$ and
$\Gamma_{a_1^\pm\to\pi^\pm\gamma}$ (\ref{a1radwid}). Variant C is  the fit including
$a_1+a_1^\prime+a_1^{\prime\prime}$ contributions. Variant D includes only $a_1+a_1^\prime$ ones.
Also shown are the corresponding calculated original \cite{bando88a,bando88} GHLS parameters
and the   magnitudes of branching fractions of the above decay.  }
\begin{ruledtabular}
\begin{tabular}{ccc}
parameter&variant C&variant D\\
\hline
$m_{a_1}$[GeV]&$1.368\pm0.006$&$1.401\pm0.006$\\
$a$&$\equiv2$&$\equiv 2$\\
$b$(calculated)&$4.89\pm0.07$&$5.37\pm0.06$\\
$c$(calculated)&$1.30\pm0.07$&$1.12\pm0.05$\\
$d$(calculated)&$-0.03\pm0.06$&$0.07\pm0.04$\\
$\alpha_4$(calculated)&$0.66\pm0.06$&$0.45\pm0.05$\\
$\alpha_5$(calculated)&$1.29\pm0.10$&$1.31\pm0.10$\\
$r$&$0.790\pm0.008$&$0.827\pm0.006$\\
$\beta$(calculated)&$1.63\pm0.12$&$1.58\pm0.12$\\
$\gamma$[GeV$^2$]&$0.31\pm0.01$&$0.35\pm0.02$\\
$m_{a^\prime_1}$[GeV]&$1.422\pm0.007$&$1.513\pm0.001$\\
$a^\prime$&$1.80\pm0.03$&$2.01\pm0.03$\\
$r^\prime$&$0.386\pm0.005$&$0.370\pm0.006$\\
$\beta^\prime$&$0.96\pm0.05$&$0.82\pm0.05$\\
$w^\prime$&$1.19\pm0.01$&$1.18\pm0.02$\\
$\psi^\prime$&$\equiv0$&$39^\circ\pm1^\circ$\\
$m_{a^{\prime\prime}_1}$[GeV]&$1.800\pm0.007$&$-$\\
$a^{\prime\prime}$&$-0.32\pm0.02$&$-$\\
$r^{\prime\prime}$&$0.36\pm0.02$&$-$\\
$\beta^{\prime\prime}$&$-0.2\pm0.2$&$-$\\
$w^{\prime\prime}$&$0.30\pm0.04$&$-$\\
$\psi^{\prime\prime}$&$10^\circ\pm8^\circ$&$-$\\
$B_{\tau^-\to\pi^+\pi^-\pi^-\nu_\tau}$&$(8.97\pm0.13)\%$&$(8.96\pm0.17)\%$\\
$\chi^2/N_{\rm d.o.f}$&45/102&95/107\\
\end{tabular}
\end{ruledtabular}
\end{table}

At last, for the sake of completeness, we give the results of the fit of the data with the single
$a_1$ resonance in the variant with $a\equiv2$ and $\beta$ fixed from
the radiative width. The fit is rather poor. To be specific,   one obtains
$m_{a_1}=1.685\pm0.006$ GeV, $r=0.973\pm0.005$,
$\gamma=0.51\pm0.02$ GeV$^2$, and $\chi^2/N_{\rm d.o.f}=389/113$. The corresponding
spectrum is shown in Fig.~\ref{fixr} with the dotted line.

\section{Discussion and conclusion}
\label{concl} ~

The simplest variant of the generalized hidden
local symmetry model with the minimal set of the counter terms in its application
to the $\tau^-\to\pi^+\pi^-\pi^-\nu_\tau$ decay is considered in the present
work. This set simultaneously solves the problem of cancelation of
the strong momentum dependence
of the $\rho\pi\pi$ vertex and provides the measured value of
the $a_1^\pm\to\pi^\pm\gamma$ decay width \cite{bando88a,bando88}. It is shown that the variant
with the single axial vector meson $a_1(1260)$ and with the above minimal set of
free GHLS parameters meets troubles when describing  the shape of the spectrum  $\pi^+\pi^-\pi^-$. No reasonable
fit can reproduce the shape, although the branching ratio $B_{\tau^-\to\pi^+\pi^-\pi^-\nu_\tau}$
agrees with the experiment.
%%%%%%%%%%%%%%%%%%%%%%%%%%%%%%%%%%%%%%%%%%%%%%%%%%%%%%%%%%%%%%%%%%%
\begin{figure}
\includegraphics[width=85mm]{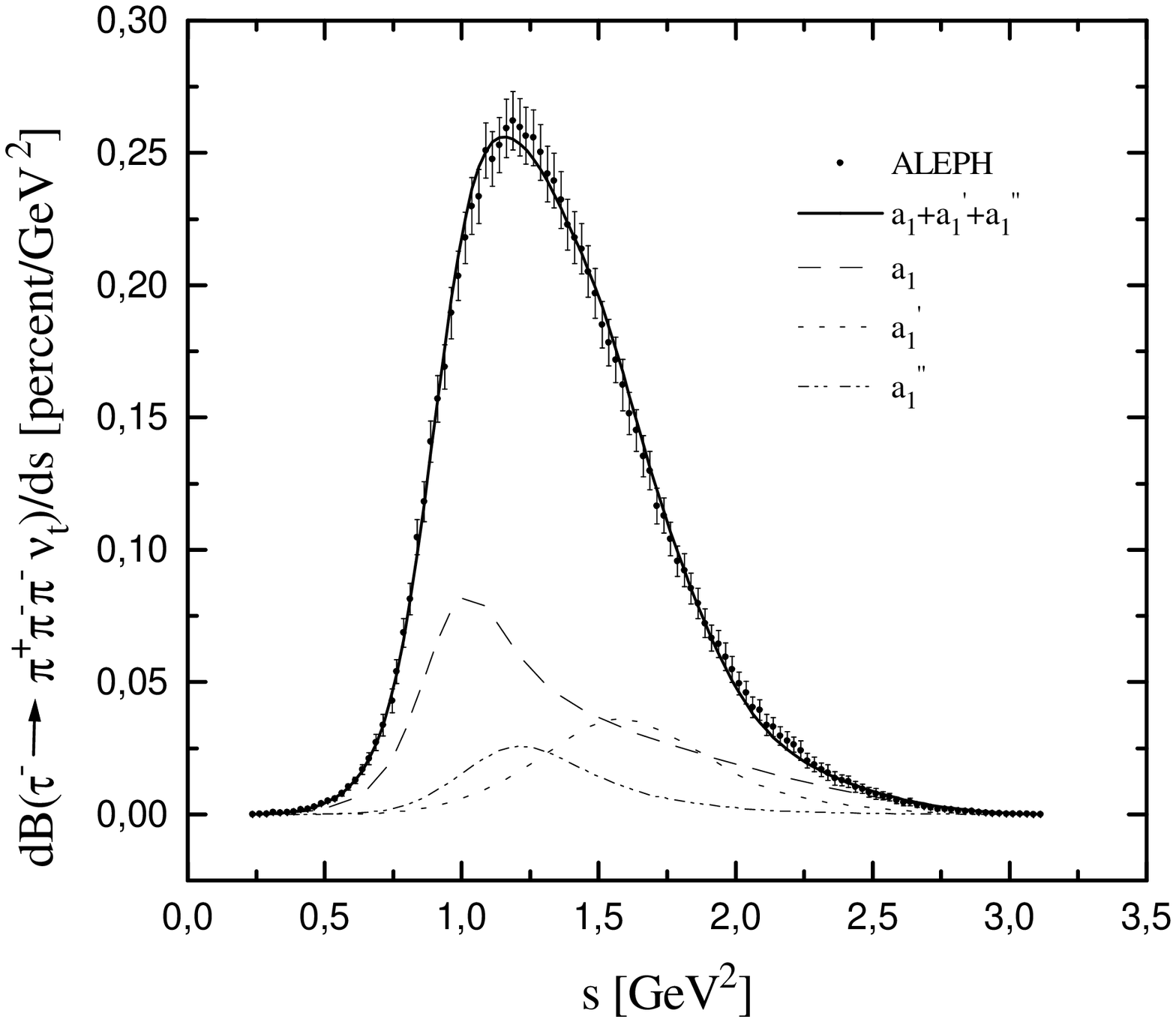}
\caption{\label{spec_const}Contributions to the spectrum of $\pi^+\pi^-\pi^-$ in $\tau$
decay in the variant A of the Table \ref{fitresults}. The dashed line
corresponds to the sum of the diagrams Fig.~\ref{taucdiag}. See the text for more  detail.}
\end{figure}
%%%%%%%%%%%%%%%%%%%%%%%%%%%%%%%%%%%%%%%%%%%%%%%%%%%%%%%%%%%%%%%%%%%%
%%%%%%%%%%%%%%%%%%%%%%%%%%%%%%%%%%%%%%%%%%%%%%%%%%%%%%%%%%%%%%%%%%%
\begin{figure}
\includegraphics[width=85mm]{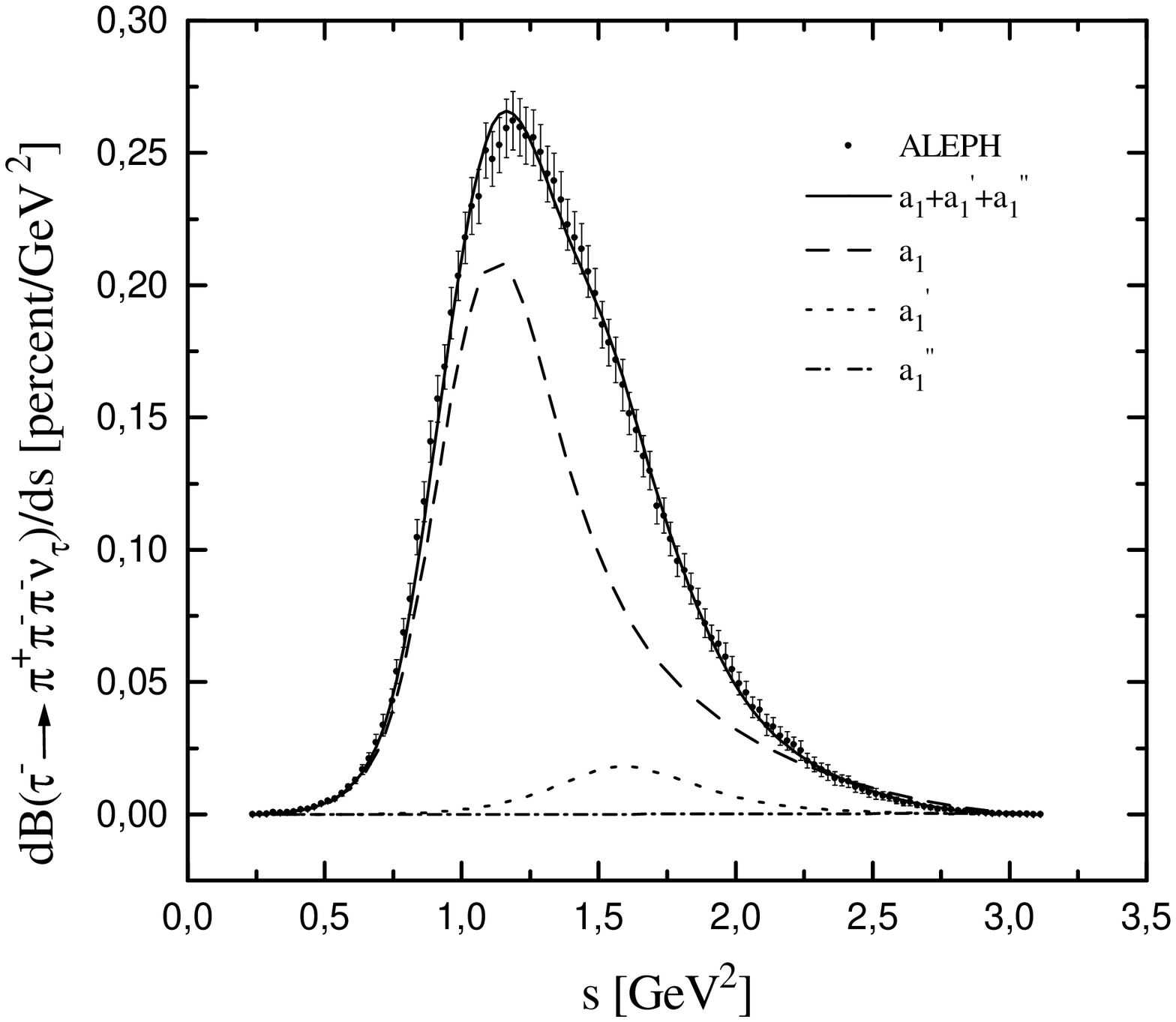}
\caption{\label{fixr_const}Contributions to the spectrum of $\pi^+\pi^-\pi^-$ in $\tau$
decay in the variant C of the Table \ref{fitresCD}. The dashed line
corresponds to the sum of the diagrams Fig.~\ref{taucdiag}. See the text for more  detail.}
\end{figure}
%%%%%%%%%%%%%%%%%%%%%%%%%%%%%%%%%%%%%%%%%%%%%%%%%%%%%%%%%%%%%%%%%%%%
One can hardly hope that higher derivatives or/and chiral loops
may improve the apparently resonant behavior. The chiral loop contributions (besides the finite width ones)
were shown to be of little importance in the four pion channel \cite{ecker}, and one cannot expect
they may help in the three pion axial vector channel, too.
As for  the higher derivative contributions, there exist the extensions \cite{li} of GHLS chiral
model with additional higher derivative terms besides the minimal set proposed in Ref.~\cite{bando88a,bando88}.
In fact, no one so far has exhausted a full list of all possible higher
derivative terms of the GHLS model. But, as we point out in first paragraph of the preceding section, any of such higher
derivative term would demand the introduction of the phenomenological form factor restricting
inevitable explosive growth of the partial width with the energy increase. This statement applies to the terms suggested in
Ref.~\cite{li} as well.

We believe that more natural is the  inclusion of the contributions of heavier axial vector mesons, in
order to reconcile calculations in GHLS with the minimal set of counter terms \cite{bando88a,bando88}
with available data \cite{aleph05}.
Like the vector resonances $\rho^\prime$, $\rho^{\prime\prime}$ \cite{pdg},
the existence of above resonances is naturally expected in the both quark model
and large $N_c$ expansion. Experiment \cite{pdg,amelin,cleo} gives   the evidence in favor of such resonances, too.
The results of the present study show that contributions of   the heavier  axial vector resonances
$a^\prime_1$ and $a^{\prime\prime}_1$  added to the $a_1(1260)$ one are capable of reproducing both
the branching ratio and
the correct shape of the pion spectrum in the decay $\tau^-\to\pi^+\pi^-\pi^-\nu_\tau$.

We would like to remind that similar problem with the simplest "canonical" GHLS
model was found
in the vector channel $e^+e^-\to\pi^+\pi^-\pi^+\pi^-$ considered earlier
\cite{ach08a,ach08b}. The  analysis presented in the above works  shows that
the contributions of heavier resonances $\rho^\prime$ and
$\rho^{\prime\prime}$ are  required for correct description of experimental data
at energy $\sqrt{s}\approx1$ GeV. However, contrary to the case of the vector channel where
additional contributions of $\rho^\prime$ and
$\rho^{\prime\prime}$ at the above energy exceed the $\rho(770)$ one,
in the axial vector channel considered in the present work, each of the additional  contributions is smaller
in magnitude  than the contribution of pure GHLS with the single fitted $a_1$ resonance.
See Fig.~\ref{spec_const} and \ref{fixr_const}. But they contribute almost coherently resulting in
the acceptable shape of the spectrum and the acceptable
magnitude of the branching fraction $B(\tau^-\to\pi^+\pi^-\pi^-\nu_\tau)\approx9\%$.
Specifically, with the central values of the fitted parameters of the variants A
in the Table \ref{fitresults} (the variants C in the Table \ref{fitresCD}), respectively,
one obtains for the net contribution of the diagrams Fig.~\ref{taucdiag}(a), (b), (c), and (d) the branching
fraction $B(\tau^-\to\pi^+\pi^-\pi^-\nu_\tau)\approx2.65(6.21)\%$. The contribution of the diagram Fig.~\ref{taucdiag}(c)
amounts to $B(\tau^-\to\pi^+\pi^-\pi^-\nu_\tau)\approx0.33(2.1)\%$. These figures should be compared with
the contributions of the diagram Fig.~\ref{taucdiag}(c) in which $a_1$ is replaced by $a^\prime_1$
and $a^{\prime\prime}_1$. One obtains $B(\tau^-\to\pi^+\pi^-\pi^-\nu_\tau)\approx1.15(0.51)\%$ and
$0.67(0.01)\%$, respectively.

%%%%%%%%%%%%%%%%%%%%%%%%%%%%%%%%%%%%%%%%%%%%%%%%%%%%%%%%%%%%%%%%%%%
\begin{figure}
\includegraphics[width=80mm]{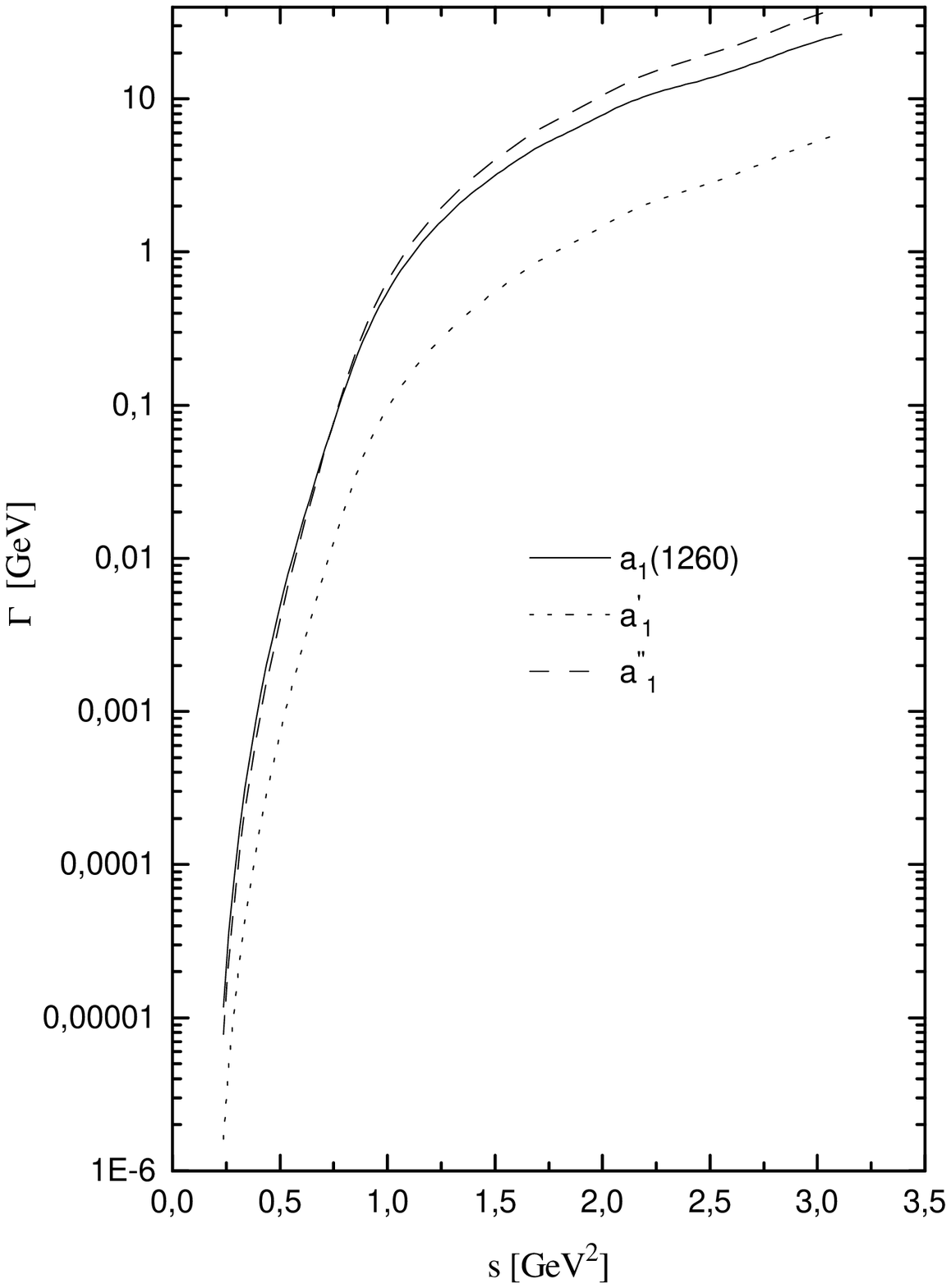}
\caption{\label{widthsa1}The widths of decay into $3\pi$
of the resonances $a_1(1260)$, $a_1^{\prime}$, and $a_1^{\prime\prime}$ evaluated with the fitted
parameters of the variant A in the Table \ref{fitresults}.}
\end{figure}
%%%%%%%%%%%%%%%%%%%%%%%%%%%%%%%%%%%%%%%%%%%%%%%%%%%%%%%%%%%%%%%%%%%%

\begin{figure}
\includegraphics[width=80mm]{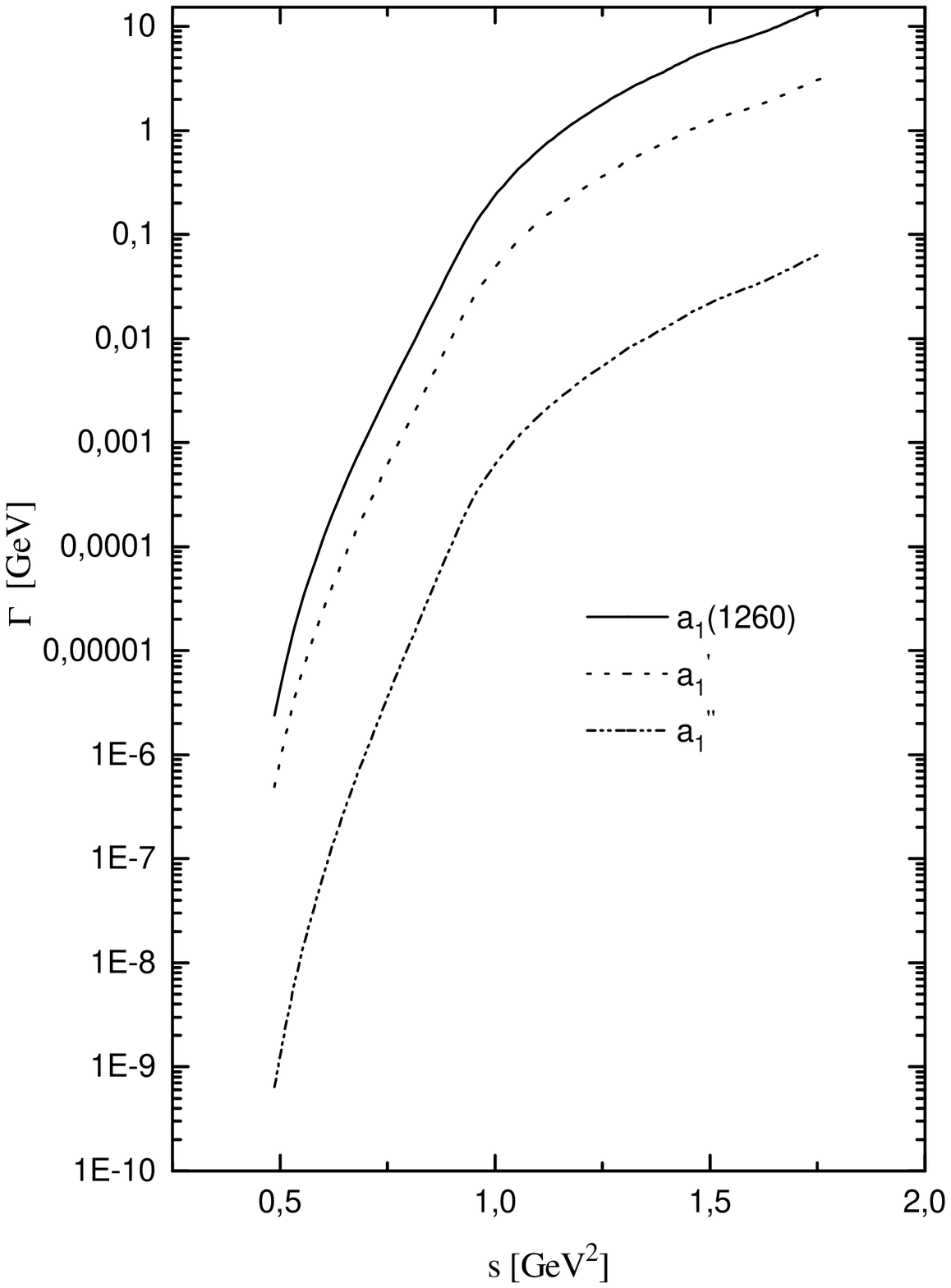}
\caption{\label{gaa1fr}The widths of decay into $3\pi$
of the resonances $a_1(1260)$, $a_1^{\prime}$, and $a_1^{\prime\prime}$ evaluated with the fitted
parameters of the variant C in the Table \ref{fitresCD}.}
\end{figure}

It is worth noting that in the variant A of the Table \ref{fitresults} the visible $a^{\prime\prime}_1$ peak
position is lower than that of $a^\prime_1$ despite of the
fact that their bare masses are in opposite relation,
see the Table \ref{fitresults}. This  can be explained as follows.
Here the dominant decay mode of $a_1^\prime$, $a_1^{\prime\prime}$
resonances  is the $3\pi$ one. Its partial width grows rapidly
with energy increase reaching the figures compatible with bare mass itself.
As was  pointed out earlier \cite{ach00a}, the combined action of the
strong energy dependence of the partial width and its large magnitude
shifts the visible peak towards the lower energies. The more the width
and the more its growth, the more the peak shifts. Since the width
of the resonance $a_1^{\prime\prime}$ and its growth are stronger as compared to
$a_1^\prime$, see Fig.~\ref{widthsa1} for the variant A of the Table \ref{fitresults},
its visible position appears
at lower energy than the visible position of $a_1^\prime$. For comparison,
the three pion decay widths of the resonances $a_1$, $a_1^\prime$, and $a_1^{\prime\prime}$
evaluated with the parameters of the variant C of the Table \ref{fitresCD}
are shown in Fig.~\ref{gaa1fr}.
%%%%%%%%%%%%%%%%%%%%%%%%%%%%%%%%%%%%%%%%%%%%%%%%%%%%%%%%%%%%%%%%%%%
%%%%%%%%%%%%%%%%%%%%%%%%%%%%%%%%%%%%%%%%%%%%%%%%%%%%%%%%%%%%%%%%%%%%

In principle, the vector isovector resonances $\rho^\prime$,
$\rho^{\prime\prime}$ could be present in the final states of
the present axial vector isovector channel $\pi^+\pi^-\pi^-$, via
the transition $a_1\to\rho^\prime(\rho^{\prime\prime})\pi\to3\pi$
(analogously for $a^\prime_1$, $a^{\prime\prime}_1$). However,
their inclusion requires the introduction of new free parameters
$g_{a_1\rho^\prime(\rho^{\prime\prime})\pi}$ (analogously for
$a^\prime_1$, $a^{\prime\prime}_1$) in addition to those 14 ones
already present.  On the grounds of reasonable
adequacy we neglect $\rho^\prime$, $\rho^{\prime\prime}$ at the
present stage of the study, especially because the coupling
constants $g_{\rho^\prime\pi\pi}$ and
$g_{\rho^{\prime\prime}\pi\pi}$ are presumably small as compared
to $g_{\rho\pi\pi}$ \cite{ach97}.

We would like to thank M.~Davier for providing us the reference to
the ALEPH data in the table form. The present work is supported in
part by Russian Foundation for Basic Research grant
RFFI-10-02-00016.

%%%%%%%%%%%%%%%%%%%%%%%%%%%%%%%%%%%%%%%%%%%%%%%%%%%%%%%%%%

\end{document}